\documentclass{article}
\usepackage{epsfig}
\usepackage{amssymb}
\usepackage{hiph-art}

\newcommand{\vb}{{\vec b}}  
\newcommand{\vp}{{\vec p}}

\newcommand{\vx}{{\vec x}} 
\newcommand{\vv}{{\vec v}}

\newcommand{\gton}{\mathrel{\lower.9ex \hbox{$\stackrel{\displaystyle 
>}{\sim}$}}}  
\newcommand{\lton}{\mathrel{\lower.9ex \hbox{$\stackrel{\displaystyle 
<}{\sim}$}}}  
\newcommand{\Tr}{{\rm Tr}}

\newcommand{\be}{\begin{equation}} 
\newcommand{\ee}{\end{equation}}  
\newcommand{\ben}{\begin{enumerate}}  
\newcommand{\een}{\end{enumerate}} 
\newcommand{\bit}{\begin{itemize}}  
\newcommand{\eit}{\end{itemize}} 
\newcommand{\bc}{\begin{center}}  
\newcommand{\ec}{\end{center}} 
\newcommand{\bea}{\begin{eqnarray}}  
\newcommand{\eea}{\end{eqnarray}}

\volnumber{19}\issuenumber{1}\edyear{2004}
\frompage{000}\topage{000}\recrevdate{\today}
\def\rightmark{Parton coalescence and spacetime}
\def\leftmark{D\'enes Moln\'ar}
 
\title{Parton coalescence and spacetime}
 
\date{\today}
 
\authors{
{D\'enes Moln\'ar$^1$ %
\index{One, D. Molnar} %
}\\[2.812mm]
{\normalsize
\hspace*{-8pt}$^1$ Department of Physics, The Ohio State University,\\
Columbus, OH 43210, USA\\[0.2ex] 
}}

\abstract{
The influence of spacetime dynamics in hadronization via 
parton coalescence at RHIC is investigated using covariant 
parton transport theory.
Key observables, the quark number scaling of elliptic flow and the 
enhancement of the $p/\pi$ ratio, show strong dynamical effects and
differ from earlier results based on the simple coalescence formulas.
}

\keyword{Relativistic heavy-ion collisions, Hadronization, Quark coalescence}

\PACS{12.38.Mh; 24.85.+p; 25.75.Gz; 25.75.-q}

\makeindex
\begin{document} 

\maketitle

\section{Introduction}
Recent exciting puzzles in $Au+Au$ reactions at $\sqrt{s_{NN}}=130$ and $200$ 
GeV at RHIC are the quark number scaling 
of elliptic flow\cite{QM2004exp,STARboth,PHENIXv2scaling}
and the weaker baryon suppression than that of mesons in 
the intermediate transverse momentum region $2 < p_\perp < 5$ GeV 
\cite{QM2004exp,STARboth,PHENIXnoBsupp,STARnoBsupp}.
Parton coalescence\cite{Voloshincoal,HwaYang,texbudMtoB,dukeCoal,coalv2,charmv2,myQM2004,QM2004th} is probably the most promising
proposal at present to explain both phenomena.

In the coalescence model,
mesons form from a quark and antiquark, 
while baryons from three quarks or three antiquarks.
The simplest version of the model is based on the ``coalescence formula''
(Eq.~(1), or minor variation of  it)
that gives the hadron spectra in terms of the 
constituent phasespace distributions on a 3D spacetime hypersurface.
Remarkably, this simple approach can  
reproduce quite well the particle spectra at RHIC\cite{texbudMtoB,dukeCoal}
and can explain the scaling of elliptic flow $v_2(p_\perp)$ with constituent
number\cite{coalv2,charmv2}.
Furthermore, 
the parameterizations of the constituent distributions and the hypersurface 
are consistent with hadronization from a thermalized 
quark-antiquark plasma at RHIC.

Nevertheless, these earlier studies left several important questions open.
For example, it is known\cite{coalv2,charmv2} that the coalescence formula
violates unitarity.
The yield in a given coalescence channel scales quadratically/cubically
with constituent number,
moreover, the same constituent contributes to several channels
(including fragmentation in certain schemes).
In addition, it is not known whether the extracted hadronization
parameters
(temperature, flow velocity, volume, hadronization time, etc.)
are consistent with any dynamical scenario.
In particular, the simple form of constituent 
phasespace distributions assumed ignores several kinds of phasespace 
correlations that would be present in a dynamical approach.

The goal of this paper is to improve upon the above deficiencies
and study how the dynamics of parton coalescence affects 
basic observables at RHIC, such as the $p$ and $\pi$ suppression pattern 
($R_{AA}$)
and elliptic flow $v_2(p_\perp)\equiv \langle \cos\phi\rangle_{p_T}$.
For simplicity, collisions at impact parameter $b=8$ fm 
($\approx 30$\% centrality) are considered.
Preliminary results were published in \cite{myQM2004}.

\section{Dynamical coalescence approach}
\label{formalism}

The parton coalescence formalism 
is largely based on studies of deuteron 
formation \cite{GFR,Dover,Nagle,Scheibl}.
In nonrelativistic $N$-body quantum mechanics, 
the number of deuterons (of given momentum) at time $t$ is 
$N_d(t) = N_{pairs} \Tr[\hat\rho_d\, \hat \rho^{(2)}(t)]$,
where $\hat \rho_d\equiv|\Phi_d\rangle\langle \Phi_d|$
is the deuteron density matrix,
$\hat \rho^{(2)}(t)$ is the projection of the density matrix 
$|\Psi^{(N)}(t)\rangle \langle \Psi^{(N)}(t)|$ of the system 
onto the (two-particle) deuteron subspace,
there are $N_{pairs}$ possible $n{-}p$ pairs,
while $\Phi_d$ and $\Psi^{(N)}(t)$ are the wave function of
the deuteron and the system.
The goal is a suitable approximation for
the observed deuteron number $N_d(t{\to}\;\infty)$
because one cannot solve the full $N$-body problem. 

One approach is to postulate that interactions {\em cease suddenly}
at some time $t_{f}$, 
and approximate $\hat \rho^{(2)}$ in the Wigner representation 
as the product of classical phase space densities $f_n \times f_p$ 
on the $t=t_{f}$ hypersurface.
The approximation, at best, is valid for weak bound states
and ignores genuine two-particle correlations.
Applied to meson formation $q\bar q\to M$, one obtains
the simple coalescence formula
\be
\frac{dN_M(\vp)}{d^3p} \! =\! g_M \! 
\int\! \prod_{i=1,2}(d^3 x_i d^3 p_i) \, W_M(\Delta \vx,\Delta \vp)\,
f_q(\vp_1,\vx_1) f_{\bar q}(\vp_2,\vx_2)\, 
\delta^3(\vp{-}\vp_1{-}\vp_2)
\label{coaleq}
\ee
with $\Delta \vx \equiv \vx_1 - \vx_2$, $\Delta \vp \equiv \vp_1 - \vp_2$,
and the meson Wigner function
$W_M(\vx,\vp) \equiv \int d^3 b\, \exp[-i\vb\, \!\vp\,]
\Phi_M^*(\vx-\vb/2) \Phi_M(\vx+\vb/2)$. The degeneracy factor $g_M$ 
takes care of quantum numbers (flavor, spin, color).
The formula for baryons involves a triple phasespace integral and
the baryon Wigner function (which depends on two relative coordinates
and two relative momenta).
The generalization to arbitrary 3D hadronization hypersurfaces is 
straightforward\cite{Dover,Scheibl}.

One difficulty with (\ref{coaleq}) is
the proper choice of the hypersurface. 
In quantum mechanics the deuteron number is constant 
at {\em any time} after freezeout, 
however for free streaming 
(\ref{coaleq}) decreases\cite{GFR,Nagle} with $t_f$.
Also, transport approaches (i.e., self-consistent freezeout) 
yield diffuse 4D freezeout 
distributions\cite{adrianFO,diffuseFO,ziweiFO},
which cannot be well approximated with a hypersurface.
These problems have been addressed in \cite{GFR}
by Gyulassy, Frankel and Remler (GFR), where they derived a way to interface
transport models and the coalescence formalism in the weak binding limit.

The GFR result is the same as (\ref{coaleq}),
except that the weight $W_M$ is evaluated using the {\em freezeout coordinates} $(t_1,\vx_1)$, $(t_2,\vx_2)$ of each constituent pair.
When taking $\Delta \vx$, the earlier particle needs to be propagated to the
time of the {\em later} one, resulting in an extra term, e.g.,
$\Delta \vx = \vx_1 - \vx_2 + (t_2 - t_1) \vv_1$ if $t_1 < t_2$.
The origin of this correction is that a weak bound state can 
only survive if none of its constituents have any further interactions.
The generalization to baryons involves propagation to the {\em latest} 
of the three freezeout times.

To investigate coalescence dynamics at RHIC,
we implanted GFR into 
covariant parton transport theory\cite{ZPC,v2,MPC,diffuseFO}.
First the parton ($g,u,d,s,\bar u, \bar d, \bar s$) evolution
was computed until freezeout via the covariant 
Molnar's Parton Cascade (MPC) algorithm\cite{MPC}.
For simplicity, only $2\to 2$ processes were considered,
with Debye-screened cross sections
$d\sigma/dt \propto 1/(t-\mu_D^2)^2$, $\mu_D = 0.71$ GeV.
For the total gluon-gluon cross section two values were explored,
$\sigma_{gg} = 3$ mb (the typical pQCD estimate) and 10 mb.
The quark-gluon and quark-quark cross sections were
suppressed by the appropriate ratio of $SU(3)$ Casimirs:
$\sigma_{qq} = (4/9) \sigma_{gq} = (4/9)^2 \sigma_{gg}$.

At parton freezeout, the GFR formula was applied (in the two- or three-body
center of mass frame) using, as common in transport approaches\cite{Nagle},
box Wigner functions
$W=\prod_{i,j} \Theta(x_m-|\vx_i-\vx_j|)\Theta(p_m-|\vp_i-\vp_j|)$,
with $x_m = 1$ fm. ($p_m$ is 
fixed by the normalization for $W$.)
This way (\ref{coaleq}) has a 
probabilistic interpretation: if $W=1$ 
(and the quantum numbers match) the hadron is formed, 
otherwise it is not ($W=0$).
If several coalescence final states existed
for a given constituent, one was chosen randomly with equal probability 
for all.
Meson channels to $\pi$, $K$, $\eta$, $\eta'$,
$\rho$, $K^*$, $\omega$, $\Phi$;
and baryon channels to
$p$, $n$, $\Sigma$, $\Lambda$, $\Xi$,
$\Delta$, $\Omega$ were considered.
Gluons were assumed to split to a $q{-}\bar q$ pair with asymmetric
momentum fractions $x\approx 0$ and $1$,
approximated as $1 g \to 1q$ (or $\bar q$).
Easy color neutralization was also assumed.
Partons that did not find a coalescence partner were fragmented 
independently via
JETSET 7.4.10 \cite{JETSET}. JETSET was also used to decay unstable hadrons.

Because (\ref{coaleq}) requires a good sampling of full 6D 
phasespace,
for the best statistical accuracy,
quarks and antiquarks were separately flavor-averaged ($(u+d+s)/3$)
at hadronization.
This assumption of flavor equilibration, which is only approximate for
strange quarks, will be lifted in the future (requires longer computation).

\begin{figure}[hbpt] 
\begin{center}
\epsfig{file=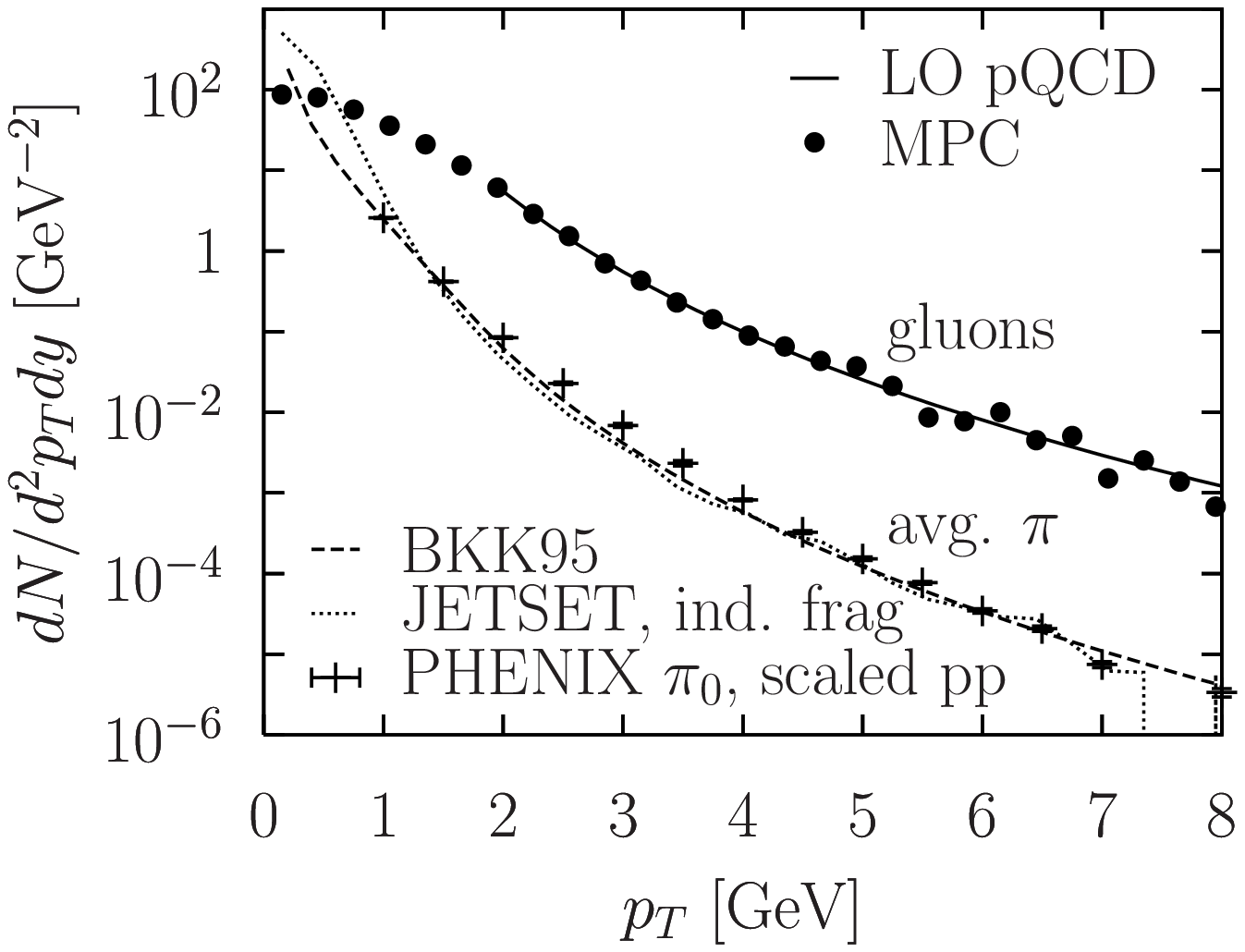,height=1.8in,width=2.5in,clip=5,angle=0}
\epsfig{file=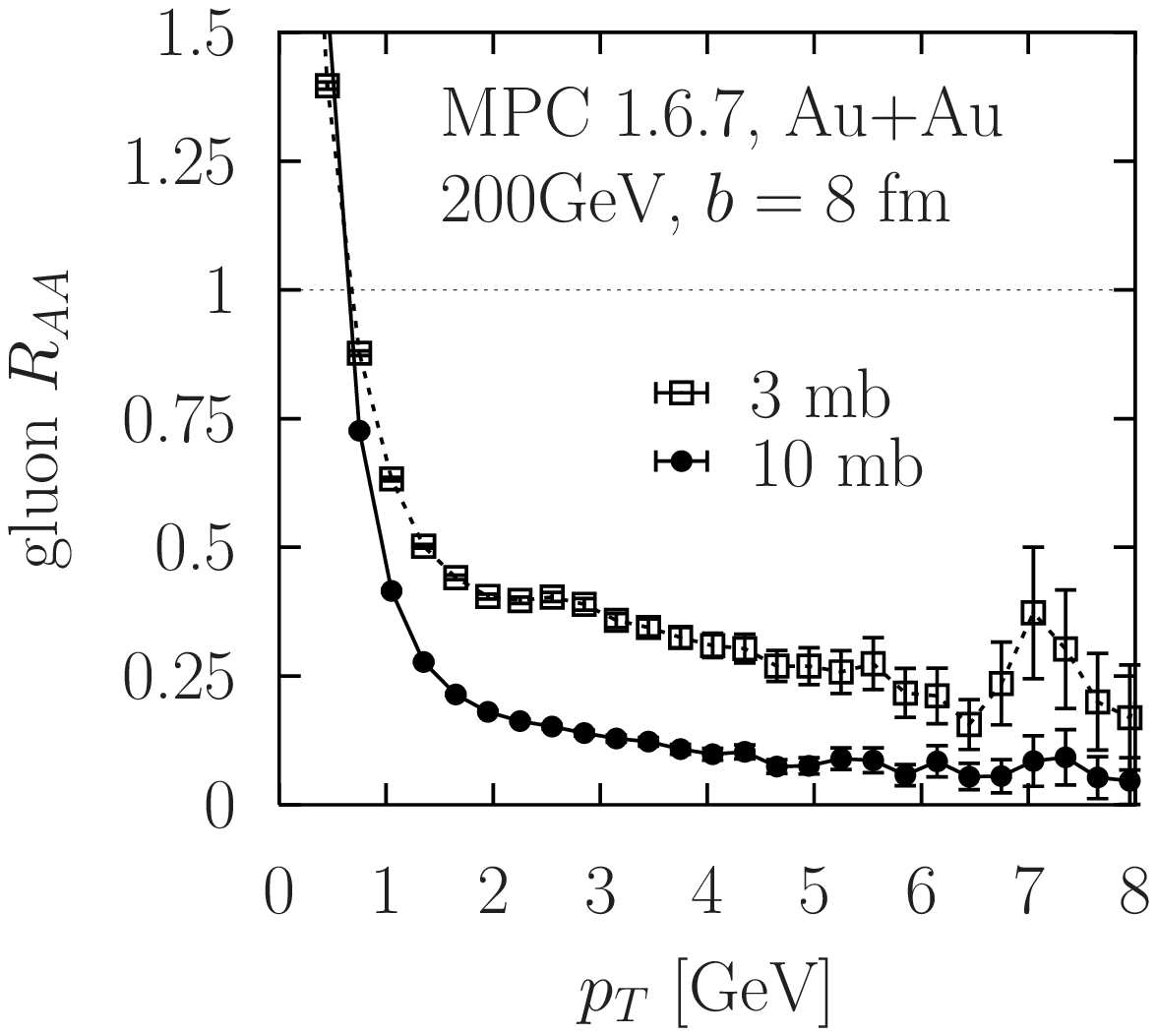,height=1.8in,width=2.3in,clip=5,angle=0}
\end{center}
\vspace*{-0.25cm}
\hskip 3.8cm a) \hskip 5.85cm b)
\vspace*{-0.4cm} 
\caption{\label{fig1}
Results for $Au+Au$ at $\sqrt{s}=200A$ GeV at RHIC with $b=8$ fm.
a) Initial gluon spectrum, and reproduction of pion spectra in $p+p$ 
collisions at RHIC (see text for details);
b) gluon quenching factor $R_{AA}$ as a function of $p_T$ for
$\sigma_{gg} = 3$ mb (open squares) and $10$ mb (solid circles).
}
\end{figure} 
The parton initial conditions for
$Au+Au$ at $\sqrt{s}=200A$ GeV at RHIC
with $b=8$ fm were taken from \cite{v2}.
However, as illustrated for {\em gluons} in Fig.~\ref{fig1}a,
for $p_\perp > 2$ GeV LO pQCD minijet three-momentum distributions 
(solid line)
were used (with a $K$-factor of 2, GRV98LO PDFs, and $Q^2{=}p_T^2$),
which below $p_\perp < 2$ GeV were smoothly extrapolated (circles) 
to yield a total parton $dN(b{=}0)/dy=2000$ at midrapidity.
This choice is motivated by the observed $dN_{ch}/dy \sim 600$ and the 
expectation that coalescence dominates the production.
Perfect $\eta=y$ correlation was assumed.

Figure~\ref{fig1}a also shows that the initial condition reproduces
fairly well
the observed pion spectra in $p+p$ at $\sqrt{s}=200$ GeV 
at RHIC. Binary collision 
scaled $\pi_0$ data from $p+p$ \cite{PHENIXpi0pp} (crosses)
are compared to the isospin-averaged pion spectrum
from the initial condition hadronized 
via either independent fragmentation in JETSET (dashed line) or 
BKK95 fragmentation functions
\cite{BKK95} (dotted line).

\section{Results on spectra and elliptic flow} 

The final hadron momentum distributions are given by a 
convolution of three dynamical effects,
i) the evolution of parton phasespace distributions due to multiple 
scatterings,
ii) the hadronization process,
iii) resonance decays (hadronic final state interactions were ignored).

The transport evolution {\em quenches} the parton spectra at 
high-$p_T$ by about a factor of three ($\sigma_{gg}=3$ mb) to ten (10 mb),
as shown in Fig.~\ref{fig1}b.
The ratio $R_{AA}$ of the parton spectrum at freezeout to the
 initial spectrum (which corresponds to no nuclear effects) is plotted.
In the language of parton energy loss models\cite{partonEloss},
the suppression in this study comes from {\em incoherent elastic} energy loss,
while in the context of hydrodynamics, it reflects
the {\em cooling} of the expanding system due to $p dV$ work.
These results are very similar to those in Ref.~\cite{v2} (see Fig.~9 therein),
even though that study considered a pure gluon gas with 
{\em thermal} initial conditions.

\begin{figure}[hbpt] 
\begin{center}
\epsfig{file=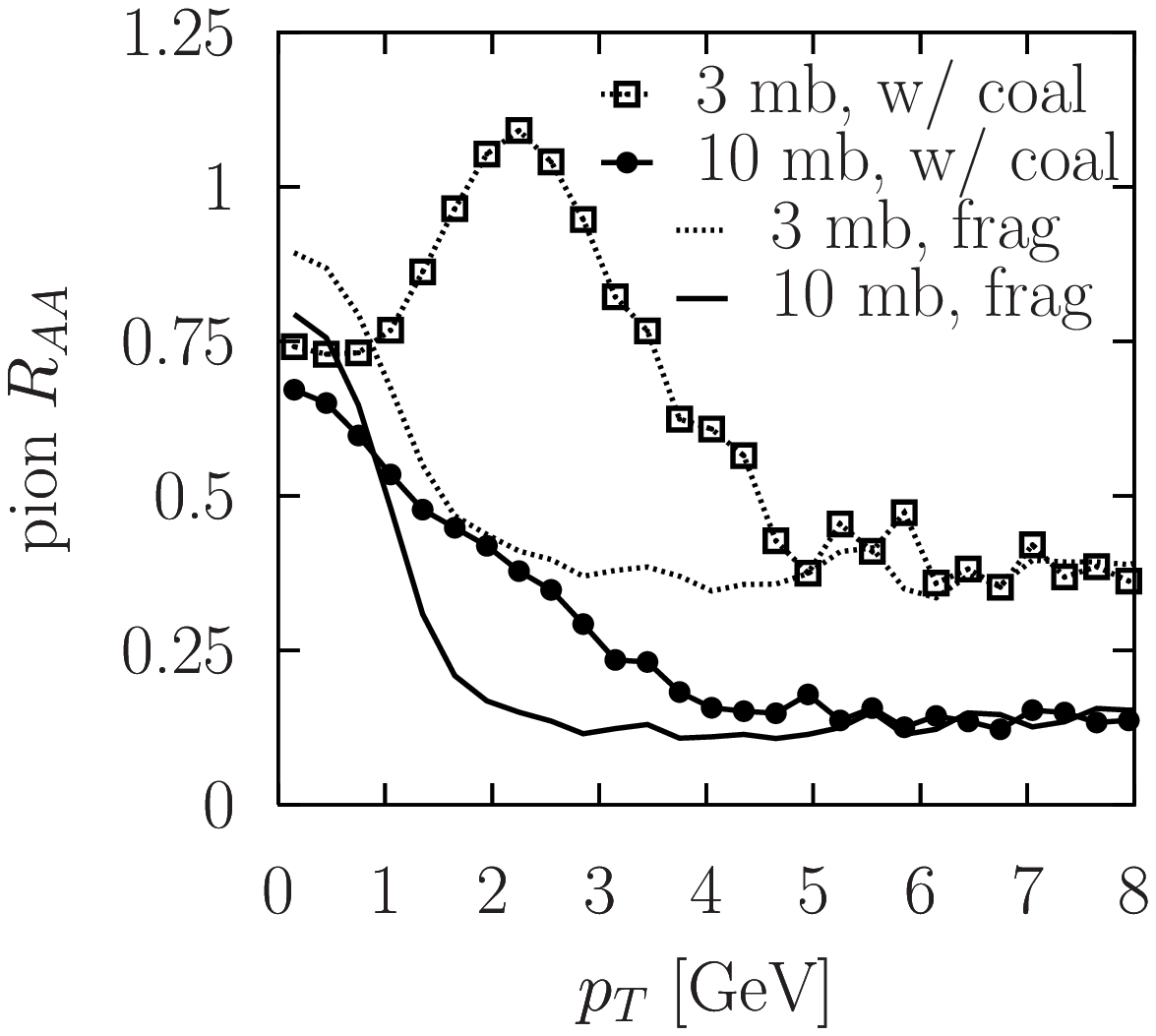,height=1.9in,width=2.45in,clip=5,angle=0}
\epsfig{file=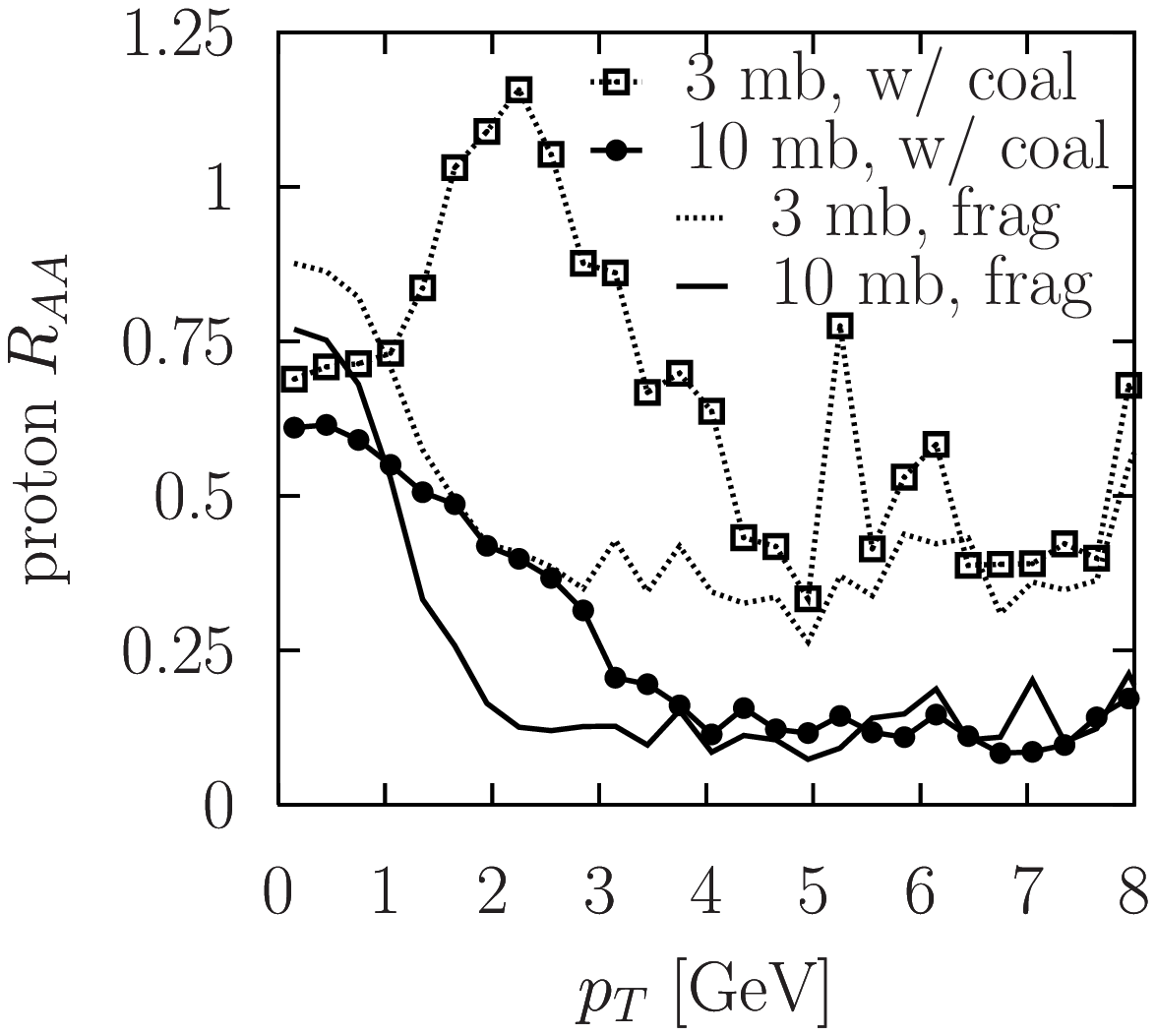,height=1.9in,width=2.45in,clip=5,angle=0}
\end{center}
\vspace*{-0.5cm} 
\caption{\label{fig2}
Pion (left) and proton (right) nuclear suppression factor at RHIC in 
$Au+Au$ at $\sqrt{s}=200A$~GeV, $b=8$ fm, with parton coalescence
(symbols) and
without coalescence (curves with no symbols), as a function of 
$p_\perp$ and parton cross section. Results for 
$\sigma_{gg} = 3$ mb (dotted) and 10 mb (solid lines) are shown.
}
\end{figure}

\begin{figure}[hbpt] 
\begin{center}
\hspace*{0cm}\epsfig{file=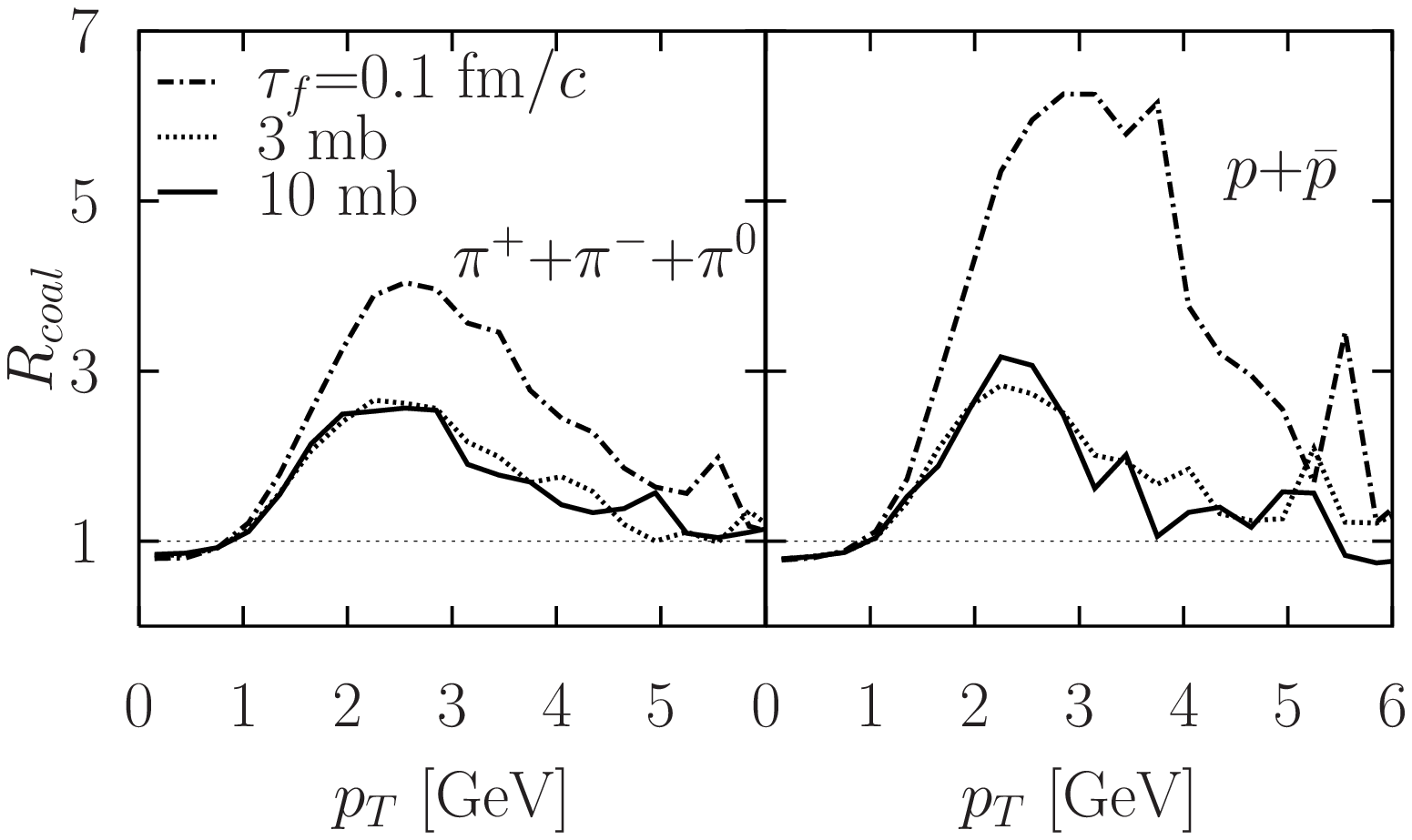,height=1.85in,width=2.95in,clip=5,angle=0}
\hspace*{0cm}\epsfig{file=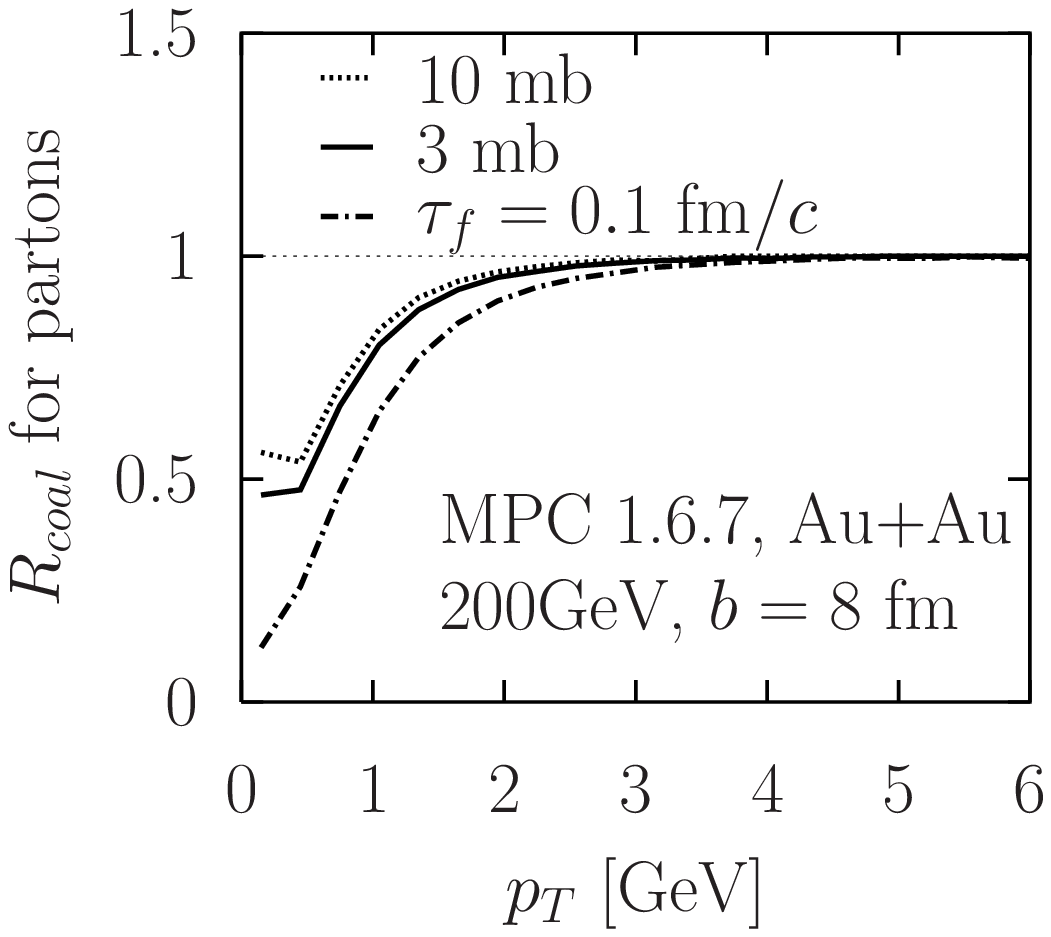,height=1.85in,width=2in,clip=5,angle=0}
\vspace*{-0.4cm} 
\end{center}
\vspace*{-0.3cm}
\hskip 4.05cm a) \hskip 6.cm b)
\vspace*{-0.5cm} 
\caption{\label{fig3}
Results for $Au+Au$ at $\sqrt{s}=200A$ GeV at RHIC with $b=8$ fm 
for $\sigma_{gg} = 3$ mb (dotted) and 10 mb (solid),
or immediate freezeout at $\tau = 0.1$ fm/c (dashed-dotted line).
a) Pion and proton enhancement from parton coalescence
as a function of $p_\perp$;
b) Fraction of partons that fragment independently as a function of $p_T$.
}
\end{figure}

Figure~\ref{fig2} shows the influence of coalescence dynamics  
on the proton and pion
nuclear suppression factor $R_{AA}$, which is the ratio
of the hadron spectra calculated from partons at freezeout
to the spectra from the initial condition hadronized via
{\em independent fragmentation}
(same as binary-scaled $p+p$).
If only fragmentation is considered, 
$R_{AA}$ for both species is about the same as that of partons 
(cf. Fig.~\ref{fig1}b).
On the other hand, parton coalescence enhances both pion and proton yields,
and hence $R_{AA}$,  by as much as a factor of three 
in the ``coalescence window'' \cite{Voloshincoal,coalv2} 
$1.5< p_T < 4.5$ GeV. The additional hadron yield comes dominantly from
partons with $0.5 < p_T < 2$ GeV, as demonstrated in Fig.~\ref{fig3}b,
where the fraction of partons that fragment independently 
(i.e., that find no coalescence partner) is plotted.
About two-thirds of the partons, including essentially all partons 
above $p_T>2.5$ GeV, hadronize via fragmentation.

Unfortunately, the inclusion of hadronization channels via coalescence does
not solve the $p/\pi$ puzzle as evident from the striking similarity between 
the left and right panels in Fig.~\ref{fig2}.
Though for mesons, the boundaries of the coalescence window agree well
with estimates\cite{texbudMtoB,dukeCoal} based on the simple 
coalescence formula (\ref{coaleq}),
for baryons the window is about the same as for mesons and 
does {\em not} extend to higher $p_T$.
Furthermore, the enhancement is about the same for pions and protons
(protons are only $\approx 5$\% higher). Therefore the $p/\pi$ ratio
at intermediate $p_T$ is the same as in $p+p$, while the data shows
an almost two-fold increase at this centrality\cite{PHENIXnoBsupp}.
In fact, the pion data\cite{PHENIXpi0AA} 
favor $\sigma_{gg} \approx 3$ mb ($R_{AA}^{\pi^0}\approx 0.4)$,
while the $p/\pi_0$ systematics\cite{PHENIXnoBsupp} 
suggests $R_{AA}^p\approx 1-1.1$,
i.e., $\sigma_{gg}\approx 10$ mb.

In Fig.~\ref{fig3}a, the enhancement is characterized by $R_{coal}$, 
the ratio of the final 
spectra with hadronization via combined coalescence and fragmentation to that 
with hadronization via fragmentation only.
Quite remarkably, $R_{coal}$ is almost independent of $\sigma_{gg}$,
despite the strong cross section dependence of the parton phasespace evolution 
shown by the quenching (cf. Fig.~\ref{fig1}b) and elliptic flow\cite{v2}.
This very interesting aspect of parton freezeout would deserve 
a more detailed study.

A qualitative explanation for the above features is that
compared to (\ref{coaleq}),
in the weak-binding case assumed,
baryon production is {\em disfavored} in a dynamical approach.
Baryons are more fragile than mesons 
because they have three constituents and therefore less chance to escape 
without further interactions.
In other words, baryons are formed at later times on average, when
the densities are smaller.
This  meson-baryon difference,
which follows from the diffuse 4D nature of self-consistent 
decoupling\cite{adrianFO,diffuseFO,ziweiFO} in spacetime,
is {\em absent} if sudden freezeout on a 3D hypersurface is postulated.
To demonstrate this, we also plot in Figs.~\ref{fig3}a-b
results for a (rather unrealistic) scenario with immediate freezeout 
on the formation 
$\tau = 0.1$fm$/c$ hypersurface, which does enhance the $p/\pi$ ratio
by a factor $1.5-1.7$
and give a wider coalescence window for baryons.

\begin{figure}[hbpt] 
\begin{center}
\epsfig{file=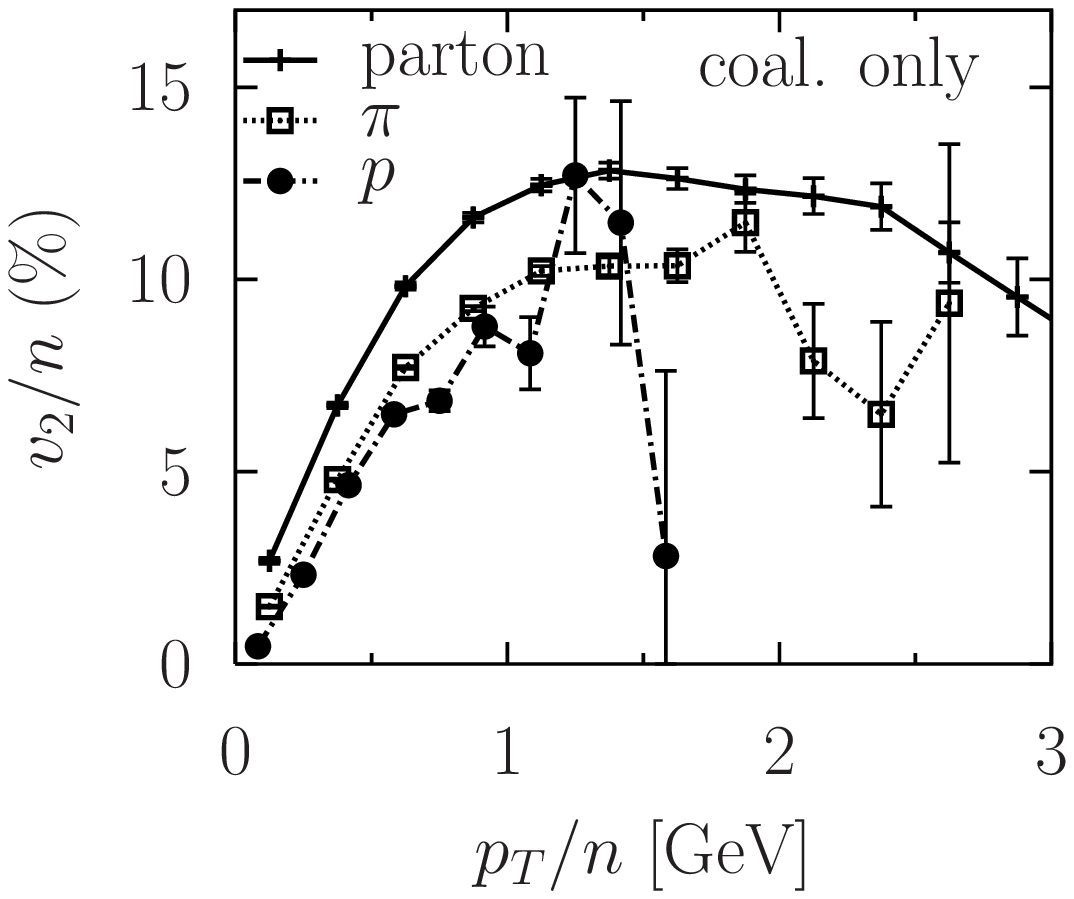,height=1.85in,width=2.45in,clip=5,angle=0}
\epsfig{file=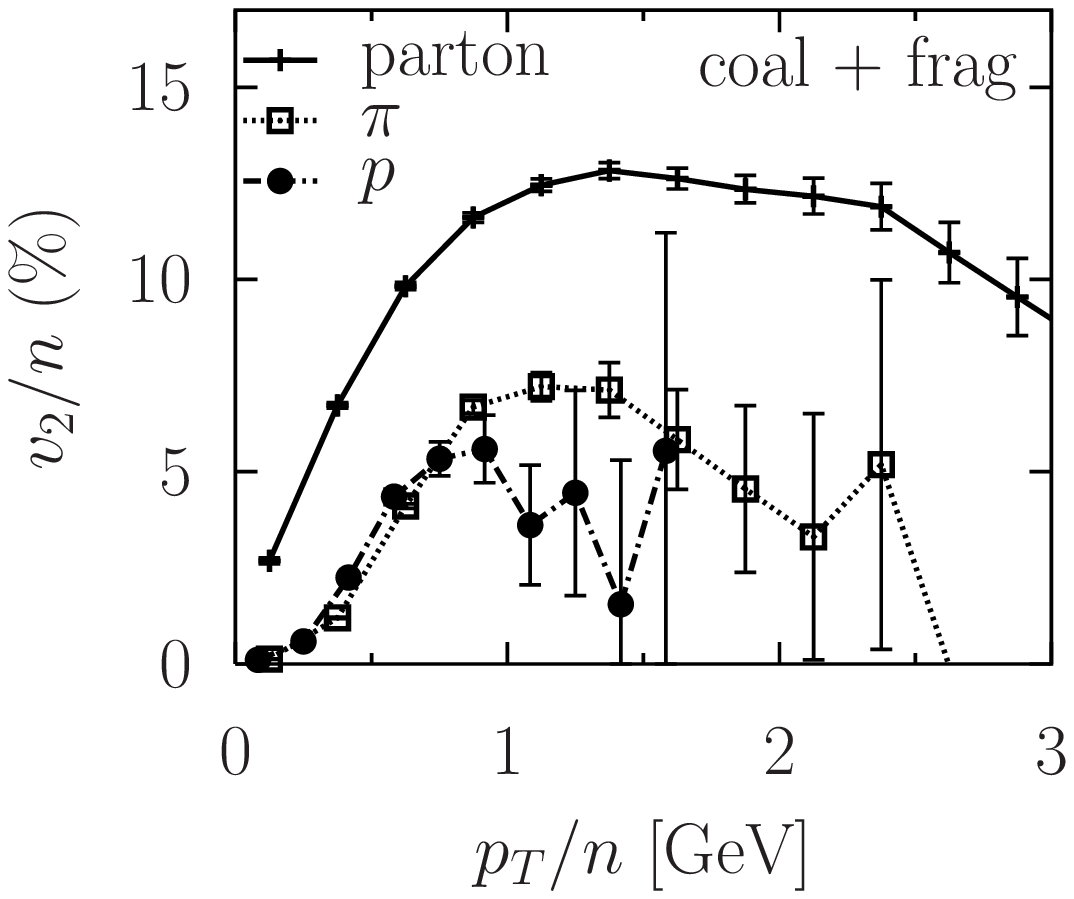,height=1.85in,width=2.45in,clip=5,angle=0}
\end{center}
\vspace*{-0.6cm} 
\caption{\label{fig4}
Quark number scaled elliptic flow as a function of $p_T$ for pions 
(open squares) and protons (filled circles)
in $Au+Au$ at $\sqrt{s}=200A$ GeV at RHIC
with $b=8$ and $\sigma_{gg}=10$ mb,
from hadronization via combined coalescence and fragmentation (right),
and for primary hadrons (i.e., without decays) from coalescence (left).
The constituent $v_2(p_T)$ is also shown (solid lines).
}
\end{figure}

Figure~\ref{fig4} shows, for $\sigma_{gg}=10$ mb,
the effect of coalescence dynamics on 
pion and proton elliptic flow, in particular
on the scaling formula\cite{Voloshincoal,coalv2}
\be
 v_2^{hadron}(p_T) = n \, v_2^{constituent}(p_T/n)
\label{v2scaling}
\ee
with constituent number $n$ ($n=2$ for mesons, 3 for baryons).
The left panel shows that the $v_2$ of {\em direct}
pions and protons from dynamical coalescence is smaller 
than predictions based on (\ref{v2scaling}),
by $20$ and $30$\%, respectively.
The reduction is much larger than 
the few-percent corrections to (\ref{v2scaling})
nonlinear in $v_2$ or those due to higher-order flow anisotropies.
It is a result of dynamical coordinate-momentum correlations that
were ignored in earlier approaches that assumed $x-p$ 
factorizable (or even spatially uniform) constituent phasespace distributions.

The right panel in Fig.~\ref{fig4} shows that
fragmentation contributions (and decays) further reduce the hadron $v_2$.
Unlike the amplification from coalescence, 
fragmentation decreases (smears out) 
the anisotropy because hadrons from the parton shower are not exactly
collinear with the originating parton 
(nonzero jet width, $\langle |\vec j_\perp| \rangle > 0$).
The parton $v_2(p_\perp)$ extracted from the hadron flows
using (\ref{v2scaling}) would underpredict the real parton $v_2$
by about a factor of two.
Also, pions and protons differ on the scaling plot
by at least 10\%, maybe even as much as 30\% 
(increased statistics to explore the $p_T/n > 1-1.5$ GeV region
is certainly desirable).
A small $10-15$\% pion-proton flow scaling violation would be allowed by
the published data \cite{STARboth,PHENIXv2scaling}.
The high-statistics Run-4 data will hopefully
provide much stronger constraints.

Despite the flow amplification due to coalescence,
the strong reduction of $v_2$ in the coalescence window,
caused  by the much smaller $\sim 25-30$\% fragmentation yield, 
``threatens'' to reopen the opacity puzzle at RHIC\cite{coalv2}.
The final magnitudes of proton and pion $v_2$
are $30-50$\% below
the data \cite{PHENIXv2scaling}. Large parton
cross sections $\sigma_{gg} \approx 20-30$ mb, an order
of magnitude above pQCD estimates, would be needed 
to generate a large enough anisotropy,
{\em at least in the approach presented here}.

\section{Conclusions}

The above findings demonstrate that coalescence is an
important hadronization channel at RHIC.
However, the results also show that dynamical effects
on the baryon/meson ratios and elliptic flow scaling 
are potentially large.

Further studies are needed to reveal
what it takes to preserve the basic 
features of the simple coalescence formulas.
It may be that the dynamics considered here was oversimplified,
or the spacetime evolution in heavy-ion collisions
is not understood well enough,
or the QCD coalescence process cannot be approximated
as a formation of weakly-bound states.

\section*{Acknowledgments}
Computer resources by the PDSF/LBNL 
and the hospitality of INT Seattle 
where part of this work was done are gratefully acknowledged.
This work was supported by DOE grant DE-FG02-01ER41190.


\begin{thebibliography}{99}

\bibitem{QM2004exp}
K.~Schweda [STAR Collaboration], nucl-ex/0403032;
M.~A.~C.~Lamont [STAR Collaboration],  nucl-ex/0403059;
J.~Castillo [STAR Collaboration], nucl-ex/0403027;
M.~Kaneta [PHENIX Collaboration],  nucl-ex/0404014.

\bibitem{STARboth}
P.~Sorensen  [STAR Collaboration],
{\it J.\ Phys.\ }G {\bf 30} (2004) S217;
J.~Adams {\it et al.}  [STAR Collaboration],
{\it Phys.\ Rev.\ Lett.\  }{\bf 92} (2004) 052302.

\bibitem{PHENIXv2scaling}
S.~S.~Adler {\it et al.}  [PHENIX Collaboration],
{\it Phys.\ Rev.\ Lett.\  }{\bf 91} (2003) 182301.

\bibitem{PHENIXnoBsupp}
S.~S.~Adler {\it et al.}  [PHENIX Collaboration],
{\it Phys.\ Rev.\ Lett.\  }{\bf 91} (2003) 172301;
Phys.\ Rev.\ C {\bf 69} (2004) 034909.


\bibitem{STARnoBsupp}
H.~Long  [STAR Collaboration]
{\it J.\ Phys.\ }G {\bf 30} (2004) S193.


\bibitem{Voloshincoal}
S.~A.~Voloshin,
{\it Nucl.\ Phys.\ }A {\bf 715} (2003) 379.


\bibitem{HwaYang}
R.~C.~Hwa and C.~B.~Yang,
{\it Phys.\ Rev.\ }C {\bf 66} (2003) 025205;
{\it ibid.} {\bf 67} (2003) 034902;
hep-ph/0312271.

\bibitem{texbudMtoB}
V.~Greco, C.~M.~Ko and P.~Levai,
{\it Phys.\ Rev.\ Lett.\  }{\bf 90} (2003) 202302;
{\it Phys.\ Rev.\ }C {\bf 68} (2003) 034904;
V.~Greco, C.~M.~Ko and R.~Rapp,
nucl-th/0312100;
V.~Greco and C.~M.~Ko,
nucl-th/0402020.


\bibitem{dukeCoal}
R.~J.~Fries {\it et al.},
{\it Phys.\ Rev.\ Lett.\  }{\bf 90} (2003) 202303;
{\it Phys.\ Rev.\ }C {\bf 68} (2003) 044902;
C.~Nonaka, R.~J.~Fries and S.~A.~Bass,
{\it Phys.\ Lett.\ }B {\bf 583} (2004) 73.




\bibitem{coalv2}
D.~Molnar and S.~A.~Voloshin,
{\it Phys.\ Rev.\ Lett.\  }{\bf 91} (2003) 092301;
D.~Molnar,
{\it J.\ Phys.}\ G {\bf 30} (2004) S235.

\bibitem{charmv2}
Z.~w.~Lin and D.~Molnar,
{\it Phys.\ Rev.\ }C {\bf 68} (2003) 044901.

\bibitem{myQM2004}
D.~Molnar, nucl-th/0403035, {\it J. Phys.} G, in press.

\bibitem{QM2004th}
J.~R.~Fries, nucl-th/0403036;
R.~C.~Hwa and C.~B.~Yang, nucl-th/0403072.


\bibitem{GFR}
M.~Gyulassy, K.~Frankel  and E.~a.~Remler,
{\it Nucl.\ Phys.\ }A {\bf 402} (1983) 596.

\bibitem{Dover}
C.~B.~Dover  {\it et al.} 
{\it Phys.\ Rev.\ }C {\bf 44} (1991) 1636.

\bibitem{Nagle}
J.~L.~Nagle {\it et al.}, 
{\it Phys.\ Rev.\ }C {\bf 53} (1996) 367.

\bibitem{Scheibl}
R.~Scheibl and U.~W.~Heinz,
{\it Phys.\ Rev.\ }C {\bf 59} (1999) 1585.


\bibitem{adrianFO}
S.~Soff, S.~A.~Bass and A.~Dumitru,
{\it Phys.\ Rev.\ Lett.\ }{\bf 86} (2001) 3981.

\bibitem{diffuseFO}
D.~Molnar and M.~Gyulassy,
{\it Phys.\ Rev.\ }C {\bf 62} (2000) 054907;
{\it Phys.\ Rev.\ Lett. }{\bf 92} (2004) 052301.

\bibitem{ziweiFO}
Z.~w.~Lin, C.~M.~Ko and S.~Pal,
{\it Phys.\ Rev.\ Lett.\ } {\bf 89} (2002) 152301.


\bibitem{ZPC}
B.~Zhang,
{\it Comput.\ Phys.\ Commun.\  }{\bf 109} (1998) 193.

\bibitem{v2}
D.~Molnar and M.~Gyulassy,
{\it Nucl.\ Phys.\ }A {\bf 697} (2002) 495;
{\it Erratum-ibid} A {\bf 703} (2002) 893.

\bibitem{MPC}
D.~Moln\'ar, MPC~1.6.7.
This parton transport code can be downloaded from the WWW at
http://www-cunuke.phys.columbia.edu/people/molnard~.

\bibitem{JETSET}
T.~Sj\"ostrand {\it et al.},
{\it Comput. Phys.\ Commun. }{\bf 135} (2001) 238.


\bibitem{PHENIXpi0pp}
S.~S.~Adler {\it et al.}  [PHENIX Collaboration],
{\it Phys.\ Rev.\ Lett.}\  {\bf 91} (2003) 241803.

\bibitem{BKK95}
J.~Binnewies, B.~A.~Kniehl and G.~Kramer,
Z.\ Phys.\ C {\bf 65} (1995) 471.

\bibitem{partonEloss}
X.~N.~Wang, M.~Gyulassy and M.~Plumer,
{\it Phys.\ Rev.}\ D {\bf 51} (1995) 3436;
R.~Baier {\it et al.},
{\it Nucl.\ Phys.}\ B {\bf 483} (1997) 291;
M.~Gyulassy, P.~Levai and I.~Vitev,
{\it Nucl.\ Phys.}\ B {\bf 571} (2000) 197;
U.~A.~Wiedemann,
{\it Nucl.\ Phys.}\ A {\bf 690} (2001) 731.



\bibitem{PHENIXpi0AA}
S.~S.~Adler {\it et al.}  [PHENIX Collaboration],
{\it Phys.\ Rev.\ Lett.}\  {\bf 91} (2003) 072301.


\end{thebibliography}
\end{document}